\documentclass[%
 reprint,
nobibnotes,
 amsmath,amssymb,
 aps,
]{revtex4-1}

\usepackage{graphicx}% Include figure files
\usepackage{dcolumn}% Align table columns on decimal point
\usepackage{bm}% bold math
\usepackage[titletoc,title]{appendix}
\usepackage{amsmath}% http://ctan.org/pkg/amsmath
\usepackage[
top    = 1.8cm,
bottom = 1.5cm,
left   = 2cm,
right  = 2cm]{geometry}

\begin{document}

%\preprint{APS/123-QED}

\title{Origin and Detection of Microstructural Clustering in Fluids\\ with Spatial-Range Competitive Interactions}

\author{Ryan B. Jadrich}
%\email{rjadrich@utexas.edu}
\thanks{Contributed equally}
\affiliation{McKetta Department of Chemical Engineering, University of Texas at Austin, Austin, Texas 78712, USA}

\author{Jonathan A. Bollinger}
%\email{jonathanabollinger@gmail.com}
\thanks{Contributed equally}
\affiliation{McKetta Department of Chemical Engineering, University of Texas at Austin, Austin, Texas 78712, USA}

\author{Keith P. Johnston}
%\email{kpj@che.utexas.edu}
\affiliation{McKetta Department of Chemical Engineering, University of Texas at Austin, Austin, Texas 78712, USA}

\author{Thomas M. Truskett}
\email{truskett@che.utexas.edu}
\affiliation{McKetta Department of Chemical Engineering, University of Texas at Austin, Austin, Texas 78712, USA}

\date{\today}

\begin{abstract}

\noindent Fluids with competing short-range attractions and long-range repulsions mimic dispersions of charge-stabilized colloids that can display equilibrium structures with intermediate range order (IRO), including particle clusters. Using simulations and analytical theory, we demonstrate how to detect cluster formation in such systems from the static structure factor and elucidate links to macrophase separation in purely attractive reference fluids. We find that clusters emerge when the thermal correlation length encoded in the IRO peak of the structure factor exceeds the characteristic lengthscale of interparticle repulsions. We also identify qualitative differences between the dynamics of systems that form amorphous versus micro-crystalline clusters.

\end{abstract}

\pacs{Valid PACS appear here}% PACS, the Physics and Astronomy
\maketitle

%\tableofcontents

% INTRODUCTION

% &
% &&&
% &&&&&
% ===============================================================================================================
\section{INTRODUCTION}
% ===============================================================================================================
% &&&&&
% &&&
% &
Complex fluids frequently possess one or more frustrating interaction lengthscales that, regardless of origin, generate micro- to mesoscale structural heterogeneity. Archetypical examples include microemulsions~\cite{Langevin1988}, block copolymers~\cite{Helfand1975,Hamley2004}, confined fluids~\cite{GohKnobler1987,Jadrich2014}, and colloidal dispersions, including proteins~\cite{Yethiraj2003,Stradner2004,PorcarLiu2010,Johnston2012,Lafitte2014,SorarufSchreiber2014}, wherein the surfactant size, block length, pore size, and screened electrostatic repulsions set the respective length scales of frustration. Despite their contextual differences, all exhibit similar transitions between homogeneous fluid states and emergent heterogeneous phases with density correlations characterized by intermediate range order (IRO), typically identified by the presence of a pre-peak at low but finite \(k\) in the static structure factor \(S(k)\)~\cite{LiuBaglioni2011}.

In the case of a pore-glass confined binary fluid system~\cite{Schemmel2003alt}, the experimental emergence of IRO has been rationalized via the behavior of the fluid thermal correlation length \(\xi_T\), which quantifies the range of correlated concentration fluctuations and the associated IRO peak width in \(S(k)\). In particular, it was demonstrated that the crossover in the temperature-density (\(T-\rho\)) plane from dispersed fluid to strong IRO corresponds to the conditions at which \(\xi_T\) reaches the pore size, i.e., the characteristic frustrating lengthscale. Such conditions enable strong, preferential segregation of the wall-attracted species from the other component which, in turn, migrates into the pore centers. Additionally, the IRO (\(T-\rho\)) crossover conditions corresponded to state points close to where the unconfined fluid reference system would otherwise exhibit liquid-liquid macrophase segregation.

Here, we extend thermal correlation length concepts to a simple model system characterized by IRO: the short-range attractive, long-range repulsive (SL) fluid, which mimics charge-stabilized colloids with van der Waals, depletion, and/or hydrophobic attractions. Various studies have demonstrated that the long-range repulsive interaction suppresses macrophase separation--which would occur for strong short-range attractions alone--in favor of IRO structures including clusters~\cite{Sciortino2004,ToledanoSciortino2009,Bomont2012,GodfrinWagnerLiu2014,ManiBolhuis2014}. However, an ongoing challenge has been to distinguish between generic IRO (i.e., presence of \emph{any} pre-peak) and clustering specifically, particularly in a way accessible to experiments~\cite{LiuBaglioni2011,GodfrinWagnerLiu2014}. One such criterion~\cite{GodfrinWagnerLiu2014} suggests that clustering emerges when the IRO peak reaches a magnitude \(S(k_{SL}^{*})\geq2.7\); this bears similarity to the empirical Hansen-Verlet single-phase rule for tracing macroscopic freezing boundaries in simple fluids~\cite{HansenVerlet1969}.

Section II presents the SL models under consideration and the simulation protocol and theoretical methodology used to characterize their behaviors. In Section III we propose a new conceptual framework and accurate criterion for clustering: namely, clusters form when the thermal correlation length \(\xi_{T}\) encoded in the IRO pre-peak of \(S(k)\) exceeds the characteristic lengthscale of the frustrating interparticle repulsive interaction. We find that this criterion also bolsters previously proposed connections between emergent IRO in SL fluids and macroscopic phase separation in corresponding reference attractive (RA) models~\cite{GodfrinWagnerLiu2014} lacking long-range repulsions. Finally, we show that the criterion makes useful predictions for fluids that form either amorphous or micro-crystalline clusters, despite striking qualitative differences in the dynamic behaviors of these two types of systems. The paper concludes in Section IV with a brief summary of our results and their relevance to experiment.

% METHODS

% &
% &&&
% &&&&&
% ===============================================================================================================
\section{METHODS}
% ===============================================================================================================
% &&&&&
% &&&
% &
Various SL interaction models are known to exhibit IRO; here we consider a canonical example 
given by the pairwise potential~\cite{Sciortino2004}
\begin{equation} \label{model:1}
\varphi_{SL}(x)\equiv 4\epsilon ( x^{-2\alpha} -x^{-\alpha} ) +A \dfrac{e^{-x/\xi_{R}}}{x/\xi_{R}}
\end{equation}
\noindent where \(x=r/d\) is a non-dimensionalized particle separation, \(d\) is the measure of particle size, \(\epsilon\) quantifies the attractive strength, and \(A\) and \(\xi_{R}\) respectively characterize the repulsion magnitude and range. We set \(\alpha=100\) in Eq.~\ref{model:1} to mimic archetypical colloids governed by core repulsions with an attraction ranges of \(O(1\%)\) of the core diameter induced via depletant effects. The long-ranged Yukawa tail mimics screened electrostatic interactions common to charge-stabilized suspensions. The corresponding RA potentials~\cite{GodfrinWagnerLiu2014} are defined by \(\varphi_{RA}(x)\equiv H(x_{0}-x)\varphi_{SL}(x)\), where \(H\) is the Heaviside step function and \(x_{0}\) is the nearest point for \(x>1\) where \(\varphi_{SL}(x)\) is zero, which eliminates the repulsive tail.

Model SL fluids defined by Eq.~\ref{model:1} can lose stability to micro-crystalline cluster phases at high attraction strengths~\cite{Sciortino2004}, in contrast to many experimental systems of interest (e.g., proteins) that do not easily crystallize. To study the latter, we also examine a simple ternary mixture of SL particles designed to frustrate crystallization. The mixture pair potentials are described by
\begin{equation} \label{model:2}
\varphi_{SL|i,j}(x_{i,j})\equiv 4[\epsilon+(1-2\delta_{i,j})\Delta_{\epsilon}] ( x_{i,j}^{-2\alpha} -x_{i,j}^{-\alpha})+A \dfrac{e^{-x_{i,j}/\xi_{R}}}{x_{i,j}/\xi_{R}}
\end{equation}
where \(\delta_{i,j}\) is the Kronecker delta, \(i,j=-1,0,1\) correspond to small, medium (\(d=1\)), and large particles respectively, \(x_{i,j}\equiv x-(1/2)(i+j)\Delta_{d}\), 
and perturbative parameter shifts to interaction size and energy, \(\Delta_{d}=0.158\) and \(\Delta_{\epsilon}=0.25\), help to thwart crystallization and promote mixing, respectively. We use systems comprising 20\% small, 60\% medium, and 20\% large particles. This combination of \(\Delta_{d}\) and composition represents a three-component approximation of 10\% polydispersity in particle size.

In examining both models, we set various combinations of the repulsive range \(\xi_{R}\) and the thermally non-dimensionalized repulsive strength \(\beta A\) (where \(\beta = 1/k_{B}T\) and \(k_{B}\) is the Boltzmann constant) while varying the non-dimensionalized attractive strength \(\beta\epsilon\). This treatment mimics systems for which the short- and long-range aspects of constituent interactions are approximately orthogonal, such as colloids with screening lengths set by particle-solvent interactions and attractions tuned via introduction of depletants~\cite{ManiBolhuis2014}.

To generate equilibrium particle configurations, we perform 3D molecular dynamics simulations of \(N=2960\) particles interacting via Eqns. 1 and 2 in the NVT ensemble with periodic boundary conditions using LAMMPS~\cite{Plimpton1995}. Due to the steepness of the repulsion, we use an integration time-step of \(0.0005\), and due to the long-range repulsion, we include interactions out to a cut-off distance of \(r_{cut}=8.0\). For all state points, the temperature is fixed at \(k_{B}T=1.0\) via a Nos\'{e}-Hoover thermostat with time-constant \(\tau=1.0\). We calculate the structure factor \(S(k)\) from simulations by numerical Fourier Transform (FT) inversion of the radial distribution function \(g(r)\). To determine whether state points are fluid, clustered, or percolating, we calculate cluster size distributions (CSDs), which quantify the probabilistic formation of \(n\)-particle aggregates, where particles are considered part of the same aggregate if their centers are within the narrow range of the attractive well. Similar to other studies~\cite{Sciortino2004,ToledanoSciortino2009,GodfrinWagnerLiu2014,ManiBolhuis2014}, 
%new 
a system is considered clustered with aggregates of preferred size \(n^{*}\) by the presence of a \emph{local maxima} in the CSD at \(n^{*}\) occurring in the range \(1 \ll n^{*} \ll N\),
%old
%a system is considered clustered with aggregates of size \(n^{*}\) by the presence of a CSD peak at \(1 \ll n^{*} \ll N\), 
and is considered percolated (at the level of the box) by a CSD peak comprised of all particles, i.e., \(n^{*} \simeq N\). 

To obtain analytical results for a broader range of potentials, we also derive theoretical thermodynamic and pair structure results via the Ornstein-Zernike (OZ) integral equation relation \(h(k)\equiv c(k)+\rho c(k)h(k)\), where \(h(k)\equiv \textup{FT}[g(r)-1]\), \(c(k)\equiv \textup{FT}[c(r)]\), \(g(r)\) is the radial distribution function, \(c(r)\) is the direct correlation function and \(\rho\) is the number density. The OZ relation is closed via the Percus-Yevick hard sphere reference, non-linear optimized random phase approximation, \(c(r)\approx \textup{exp}[-\beta \varphi(r)]-1+G(r)\), where \(G(r)=0\) for \(r>d\) while for \(r\leq d\) it is optimized to enforce \(h(r)=-1\) (thus, we approximate Eqn. 1 with a literal hard core for \(r\leq d\))~\cite{HansenMcDonald2006}. In carrying out these calculations, we consider only the Eqn. 1 potential since non-crystalline states are avoided due to the enforcement of homogeneity. This closure yields a spinodal locus at all densities, an important feature for the RA cases.

% RESULTS & DISCUSSION

% &
% &&&
% &&&&&
% ===============================================================================================================
\section{RESULTS AND DISCUSSION}
% ===============================================================================================================
% &&&&&
% &&&
% &

\begin{figure}
  \includegraphics{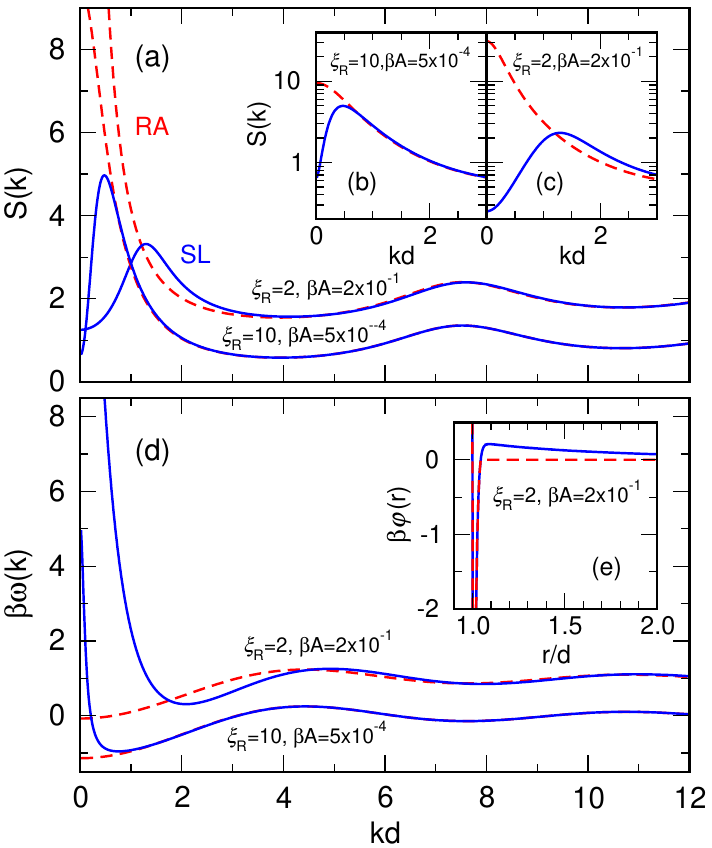}
  \caption{(Color online). (a) Structure factors \(S(k)\) for reference attractive (RA, red dashed) and short-range attractive long-range repulsive (SL, blue solid) fluids at packing fraction \(\phi=0.125\) for repulsions with ranges \(\xi_{R}\) and strengths \(\beta A\). Curves are derived from integral equation theory, where the \(\xi_{R}=10\) curves are shown for attraction \(\beta\epsilon=4.35\) and the \(\xi_{R}=2\) curves (shifted vertically) are shown for \(\beta\epsilon=4.75\). (b,c) \(S(k)\) curves from (a) replotted to highlight \(k\rightarrow0\) behaviors. (d) Fourier transforms \(\beta\omega(k)\) of the potentials from (a) with \(\xi_{R}=2\) curves shifted vertically. (e) RA and SL potentials \(\beta\varphi(r)\) for the \(\xi_{R}=2\) case.}
  \label{sch:Figure1}
\end{figure}

To begin our discussion, we first consider the behavior of the structure factor \(S(k)\) for SL fluids with different relative (integrated) repulsive strengths and corresponding RA systems (see Fig.~1a-c) as predicted from integral equation theory. The two SL fluids exhibit pre-peaks characteristic of IRO at wavelengths \(k^{*}_{SL}>0\), indicating preferential structuring on microscopic lengthscales of \(2\pi/k^{*}_{SL}\approx12.6d\) and \(5.0d\), respectively. In contrast, for the RA fluids lacking long-range repulsions, the short-range attractions drive ordering on the macroscopic lengthscale, corresponding to the peak at \(k_{RA}^{*}=0\). Crucially, we see that for the very weak repulsive case (\(\xi_{R}=10\), \(\beta A=5\times 10^{-4}\)) , the \(S(k)\) for the SL fluid traces the RA curve down to low-\(k\), supporting the conceptual notion of SL fluids as perturbations to underlying RA fluids for which only the principal ordering lengthscale has been shifted.

To understand why one should naturally expect SL fluids to aggregate on smaller lengthscales than their RA counterparts, we examine in Fig. 1(d) the Fourier space analogs of the SL and RA pair potentials, \(\omega(k) = \textup{FT}[\varphi_{0}(r)]\), where \(\varphi_{0}(r)=H(r-d)\varphi(r)\). Viewing the potentials in this way makes explicit the idea that structural oscillations of different lengthscales are \emph{weighted} by the energy profile \(\omega(k)\), which is evidenced by the close reciprocal correspondence between basins in \(\omega(k)\) (Fig. 1(d)) and peaks in \(S(k)\) (Fig. 1(a)). This connection can also be made more formal by considering microstate configurational energies (see Appendix A).

\begin{figure}
  \includegraphics{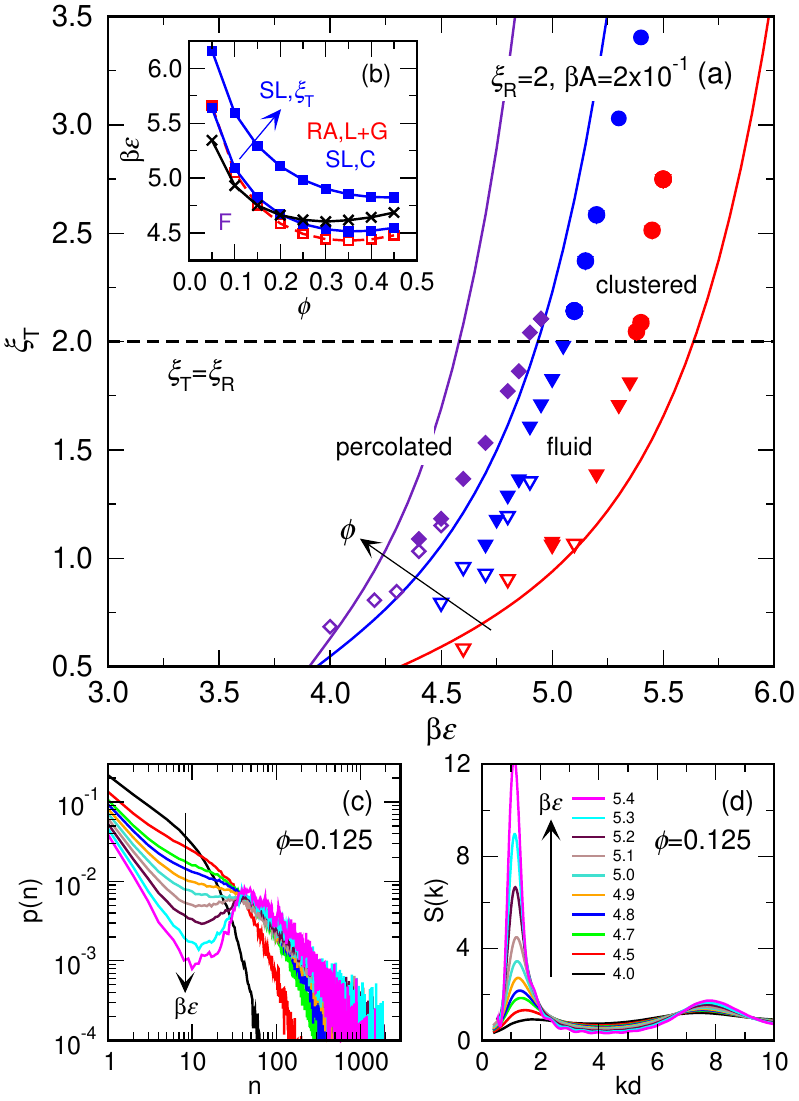}
  \caption{(Color online). (a) Symbols show thermal correlation lengths \(\xi_{T}\) for SL simulations of polydisperse (filled) and monodisperse (unfilled) systems with attractive strengths \(\beta\epsilon\) and packing fractions \(\phi =\) 0.050, 0.125, and 0.250. Symbol shapes indicate whether the state point is dispersed fluid (triangles), clustered (circle), or percolated (diamond), and the horizontal dashed line indicates \(\xi_{T}=\xi_{R}\). Solid lines show \(\xi_{T}\) calculated via theory. (b) Phase behavior calculated via theory for potentials from (a), including RA macrophase spinodal (red unfilled squares); SL curves (blue filled squares) corresponding to \(\xi_{T}=2\) and \(\xi_{T}=5\); and \(S(k^{*}_{SL})=2.7\) curve (black x). `L+G' indicates liquid-gas coexistence, `C' indicates clustered phase, and `F' indicates fluid phase. (c) Cluster size distributions indicating probability \(p(n)\) of \(n\)-particle cluster formation and (d) \(S(k)\) profiles from polydisperse simulations at \(\phi=0.125\).}
  \label{sch:Figure2}
\end{figure}

Moving beyond the above discussion concerning \emph{generic} IRO, we demonstrate in Fig. 2 that particle \emph{clustering} emerges when the thermal correlation length \(\xi_{T}\) surpasses the characteristic lengthscale of interparticle repulsion \(\xi_{R}\). Here, we estimate \(\xi_{T}\) from the well-known \(S(k)\) approximation (inverse expansion) near \(k^{*}\)~\cite{HansenMcDonald2006}:
\begin{equation} \label{theory:2}
S(k)\equiv \dfrac{S(k^{*})}{1+(k-k^{*})^2 d^2 \xi_{T}^2} 
\end{equation}
% revised
That \(\xi_{T}\) is a correlation length is evident by considering the real-space form of Eqn. 3, \(\lim_{r \to \infty}[g(r)-1] \propto r^{-1}\exp[-r/d\xi_{T}]\cos[rk^{*}-\theta]\), where \(g(r)\) is the radial distribution function, \(\theta\) is a constant, and \(\xi_{T}\) gives the characteristic decay-length of static correlations while the cosine term reflects modulated structure. In practice, \(\xi_{T}\) can be extracted from \(S(k)\) by fitting \(S(k^*)/S(k)\) to the form \(1+(k-k^{*})^2 d^{2}\xi_{T}^2\) about \(k^*\).

In Fig. 2(a), we catalog the phase behavior as a function of attractive strength \(\beta\epsilon\) for various packing fractions \(\phi\). It is evident that for the lower-density isochores, the \(\xi_{T} \geq \xi_{R}\) criterion demarcates when clustering begins in our polydisperse system, as indicated by a characteristic CSD peak with increasing attractions (Fig. 2(c)) and reflected by a growing IRO pre-peak in \(S(k)\) (Fig. 2(d)). As is intuitively expected and seen by others~\cite{ToledanoSciortino2009,ValadezLiu2013,GodfrinWagnerLiu2014}, for denser isochores like \(\phi=0.250\), it is challenging to identify precisely when ``clustering'' begins because the CSD indicates box-wide percolation (geometrically merged clusters) even down to relatively low \(\beta\epsilon\). Fig. 2(a) also shows that correlation lengths of monodisperse and polydisperse systems coincide upon approach to the \(\xi_{T}=\xi_{R}\) threshold, where this boundary also approximately identifies where the monodisperse fluid loses stability with respect to formation of micro-crystalline clusters.

In Fig. 2(b), we also examine phase behaviors for the SL and RA fluids derived for a wider \(\beta\epsilon - \phi\) parameter space via theory, which reveals close correspondence between the SL \(\xi_{T}=\xi_{R}\) boundary and the spinodal associated with RA macrophase separation. Their similar shapes (and, in this case, locations), suggest that the SL \(\xi_{T}=\xi_{R}\) boundary echoes the RA thermodynamic instability, where the frustrating repulsion has erased (or highly suppressed) liquid-gas coexistence in favor of clustering. (We also include the \(\xi_{T}=5\) curve to demonstrate the general propagation of the RA spinodal shape with increasing \(\beta\epsilon\).) As a further comparison, the empirical clustering condition \(S(k_{SL}^{*})\geq2.7\) is also shown. While it lies within similar proximity to the RA spinodal, it possesses a noticeably different, shallower contour.

\begin{figure}
  \includegraphics{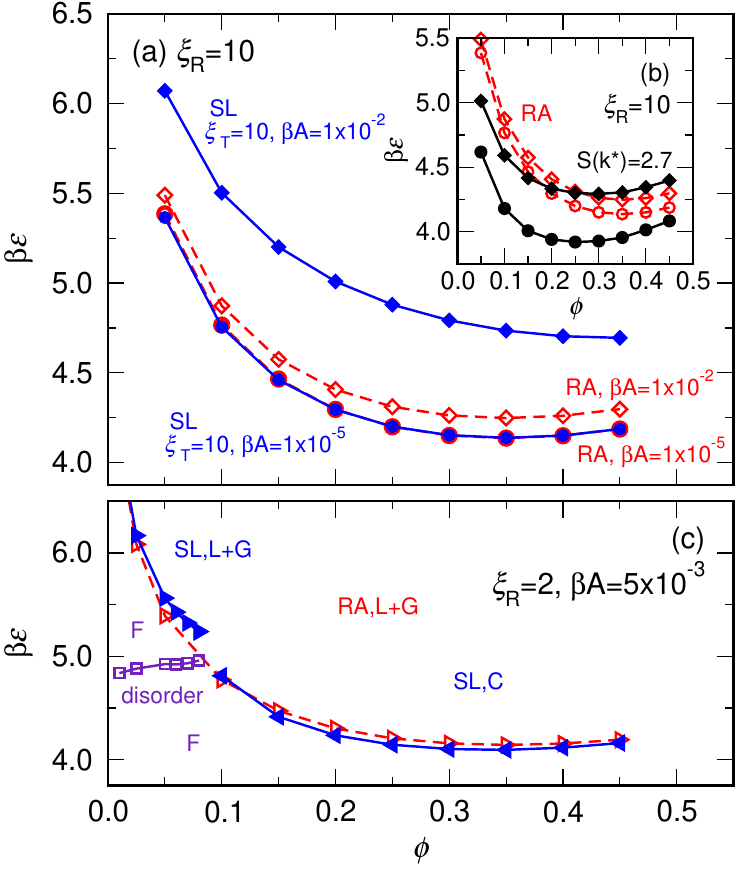}
  \caption{(Color online). (a) Phase diagrams calculated via theory, comprising RA macrophase spinodals (unfilled red symbols) and SL \(\xi_{T}=\xi_{R}\) curves (filled blue symbols) for \(\xi_{R}=10\) and two repulsive strengths \(\beta A\). (b) RA spinodals and curves along which \(S(k^{*}_{SL})=2.7\) (filled black symbols) for same systems as in (a). (c) Phase diagram calculated via theory comprising RA macrophase spinodal (unfilled red triangles); SL curves corresponding to macrophase spinodal at low \(\phi\) (right-pointing blue triangles) and \(\xi_T=\xi_R=2\) at high \(\phi\) (left-pointing blue triangles); and disorder line (purple squares) in the fluid region (see text). `L+G' indicates liquid-gas coexistence, `C' indicates clustered phase, and `F' indicates fluid phase. 
}
  \label{sch:Figure3}
\end{figure}

To elucidate deeper connections between the contours in Fig. 2(b), we explore in Fig. 3 whether the  \(\xi_{T}=\xi_{R}\) and RA spinodal boundaries truly converge for ultra-weak repulsions, which might be expected if the latter can be considered a natural weak-repulsion \emph{limit} of the former. In Fig. 3(a-b), we examine two potentials with different repulsive strengths: for \(\beta A=1\times 10^{-2}\), the repulsion is evidently ``strong'' and there is no overlap between the \(\xi_{T}=\xi_{R}\) and RA spinodal boundaries (note: this highlights that these boundaries do not generally overlap as in Fig. 2(b)). However, as repulsion strength is lowered to \(\beta A\leq1\times 10^{-5}\), the two curves collapse and become truly indistinguishable, reflecting a deep SL-RA connection. In Fig. 3(b), we also show corresponding \(S(k_{SL}^*)=2.7\) curves. Clear discrepancies in shape are apparent when comparing the RA spinodals and the \(S(k_{SL}^{*})=2.7\) boundaries, and the two types of curves increasingly move apart as \(\beta A\) is reduced.

To further generalize the connection of the RA spinodal to the phase behaviors of SL systems, we consider in Fig. 3(c) a less long-ranged weak repulsion (\(\xi_R=2, \beta A=5\times 10^{-3}\)), which exhibits intriguing properties: a true SL spinodal separation occurs for \(\phi\leq0.09\), while for higher volume fractions there is a \(\xi_{T}=\xi_{R}\) clustering boundary. The low-density fluid also exhibits a disorder line, below which the IRO peak is present and above which the IRO peak transitions to a \(k_{SL}^*=0\) peak. The intimate correspondence between the SL boundaries and the RA spinodal further reflects that the condition \(\xi_{T}=\xi_{R}\) reflects a muted thermodynamic instability, which for very weak repulsions can also emerge within the SL fluid itself.

\begin{figure}
  \includegraphics{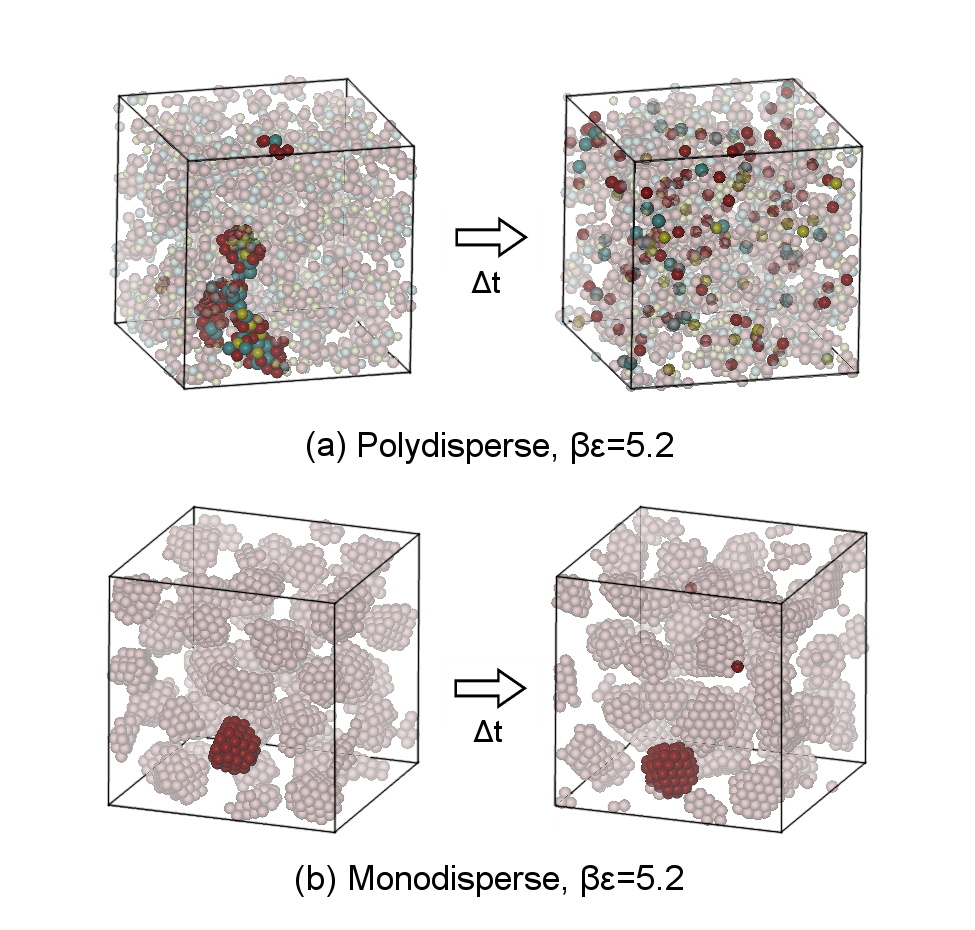}
  \caption{(Color online). Cluster phase simulation snapshots of polydisperse (a) and monodisperse (b) systems at \(\phi=0.125\) with attractive strength \(\beta\epsilon=5.2\) and repulsions defined by \(\xi_{R}=2\) and \(\beta A=0.20\). Particles comprising a single cluster (determined at time \(t\)) are rendered opaque in their positions at times \(t\) (left) and \(t'=t+\Delta t\) (right). The lag time is \(\Delta t = 25\tau_{d}\), where \(\tau_{d} = d^2/D\) is the characteristic time for \(d=1\) particles to diffuse and \(D\) is the long-time bulk diffusion coefficient determined via mean-squared displacements. Colors correspond to small, medium (\(d=1\)), and large particles, which are shaded yellow, red, and blue, respectively. Visualizations created with VMD~\cite{Humphrey1996}.}
  \label{sch:Figure4}
\end{figure}

Finally, we consider the morphologies and lifetimes of the clusters that form in polydisperse and monodisperse SL systems. Clusters in the former exhibit amorphous and irregular shapes, as exemplified by the simulation snapshots in Fig. 4(a), which correspond to the system in Fig. 2 at conditions slightly above the clustering transition. Here, it is evident based on the time-lag snapshots that the clusters are transient and continuously redistribute particles to create new clusters at the expense of others. By significantly increasing the attractive strength \(\beta\epsilon\), one can eventually observe arrested, percolating, amorphous gels as exemplified by the simulation snapshots for \(\phi=0.125\) systems in Fig. 5. Interestingly, our model gels may be thermoreversible with no local crystallinity, possibly providing a simpler alternative to valence-limited gel-formers~\cite{Sciortino2011}. Thermoreversibility is highly desired to facilitate fabrication of massively reconfigurable, reversible materials.

In contrast, monodisperse systems at similar attraction strengths can undergo highly regular clustering via local crystallization, as exemplified in Figure 4(b). While the crystalline nature of such simulated clusters has been observed previously by others~\cite{Sciortino2004,ToledanoSciortino2009,ManiBolhuis2014}, we do note that the relatively weaker repulsion examined here drives formation of much larger clusters that are more obviously crystalline in nature. The crystalline clusters are relatively static objects once formed, as demonstrated by the time-lag snapshots, in direct contrast to the amorphous clusters.

\begin{figure}[htp]
  \includegraphics{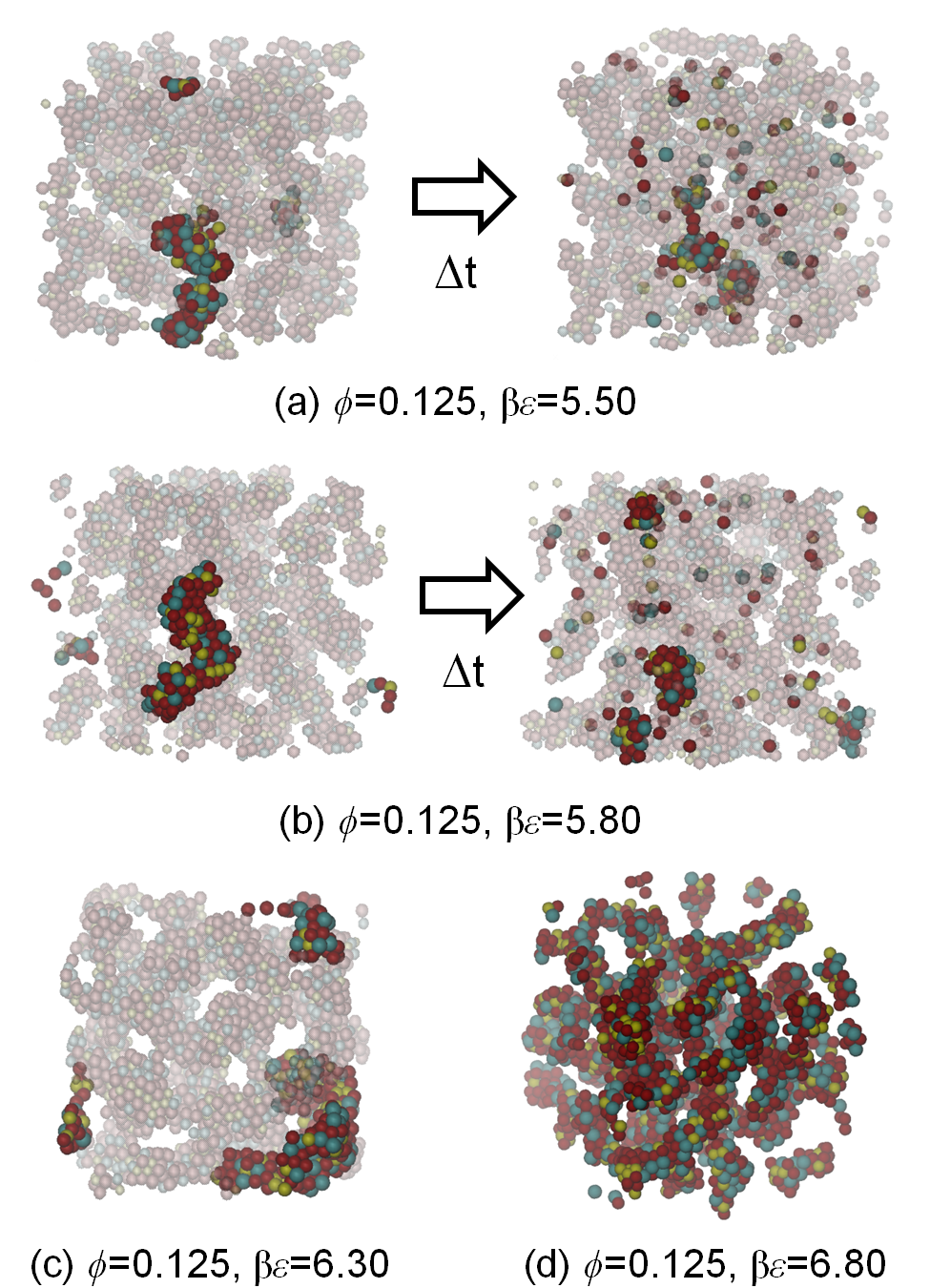}
  \caption{\linespread{1.0}\selectfont{} (Color online). Cluster phase simulation snapshots of polydisperse systems at \(\phi=0.125\) with various attractive strengths \(\beta\epsilon\) and repulsions defined by \(\xi_{R}=2\) and \(\beta A=0.20\). %, where it is clear that as \(\beta\epsilon\) is increased far above the \(\xi_T=\xi_R\) boundary, arrested percolating states resembling gels are eventually formed. 
In all snapshots, particles comprising a single cluster (determined at time \(t\)) are rendered opaque in their positions at time \(t\). For cases (a) and (b) that are not gelled, the same particles are also shown in their positions at \(t'=t+\Delta t\). The lag time in (a) and (b) \(\Delta t = 25\tau_{d}\), where \(\tau_{d} = d^2/D\) is the characteristic time for \(d=1\) particles to diffuse and \(D\) is the long-time bulk diffusion coefficient determined via mean-squared displacements. For cases (c) and (d), the configurations are dynamically arrested and \(\tau_{d}\) cannot be practically measured within the timescale of simulations. Colors correspond to small, medium (\(d=1\)), and large particles, which are shaded yellow, red, and blue, respectively. Visualizations created with VMD~\cite{Humphrey1996}.}
  \label{sch:Figure5}
\end{figure}

% &
% &&&
% &&&&&
% ===============================================================================================================
\section{CONCLUDING REMARKS}
% ===============================================================================================================
% &&&&&
% &&&
% &

In closing, we have presented a new framework for understanding and detecting cluster phases in SL fluids based on the thermal correlation length \(\xi_{T}\). This framework should prove useful for probing micro-structural transitions in diverse systems governed by frustrated interactions, e.g., lattice spin models with opposing nearest-neighbor and higher-order couplings. We have also presented the first non-microcrystallizing SL fluid, which exhibits amorphous transient clusters; this should prove useful for examining the (zeroth order) physics of real dispersions known to be resistant to crystallization, e.g., proteins. 

Finally, we remark that the \(\xi_{T}=\xi_{R}\) clustering criterion can be implemented in experiments provided that, in addition to extracting \(\xi_{T}\) from an \(S(k)\) profile (described earlier), one can also obtain a reasonable measure of the repulsive lengthscale between particles \(\xi_{R}\). For systems accurately described by simple screening models, \(\xi_{R}\) can be directly estimated. Otherwise, one can first obtain the \(r\)-space total correlation function \(h(r)\) via an inverse FT of \(S(k)\). Likewise, one can calculate the direct correlation function \(c(k)=\rho^{-1}-[\rho S(k)]^{-1}\) and then obtain its \(r\)-space equivalent \(c(r)=\textup{FT}^{-1}[c(k)]\), which provides information about the interparticle interactions because \(\lim_{r \to \infty}c(r)\approx \varphi(r)\)~\cite{HansenMcDonald2006}. By plotting \(\ln\{|r h(r)|\}\) and \(\ln\{|r c(r)|\}\) versus \(r\) (where \(|x|\) is the absolute value of \(x\)) and comparing their (negative) slopes, one directly compares the range of interparticle correlations (as captured by \(\xi_{T}\)) and the characteristic range of the interparticle interactions, respectively. Thus, given an \(S(k)\) profile exhibiting an IRO peak, if \(\ln\{|r h(r)|\}\) decays \emph{more slowly} than \(\ln\{|r c(r)|\}\), then the \(\xi_{T}\) associated with IRO exceeds the characteristic (repulsive) lengthscale \(\xi_{R}\).

% orig.
%Otherwise, one can in principle obtain \(\xi_{R}\) from \(S(k)\) by calculating \(c(k)=\rho^{-1}-[\rho S(k)]^{-1}\) and obtaining its \(r\)-space equivalent \(c(r)=\textup{FT}^{-1}[c(k)]\), where \(\lim_{r \to \infty}c(r)\approx \varphi(r)\)~\cite{HansenMcDonald2006}. Thus, given knowledge of the long-range behavior of \(\varphi(r)\), one can recover the long-range decay lengthscale \(\xi_R\). 

% ACKNOWLEDGMENTS

\begin{acknowledgments}
This work was partially supported by the National Science Foundation (1247945), the Welch Foundation (F-1696), and the Gulf of Mexico Research Initiative. We acknowledge the Texas Advanced Computing Center (TACC) at The University of Texas at Austin for providing HPC resources.
\end{acknowledgments}

% &
% &&&
% &&&&&
% ===============================================================================================================

\section*{APPENDIX A: FOURIER-SPACE CONNECTIONS BETWEEN \(\omega(k)\) AND \(S(k)\)}

An \(N\) particle configuration [\(\textbf{r}_{i}\)] that does not violate the hard core constraint is weighted according to the Boltzmann factor \(\textup{exp}[-\beta \Omega([\textbf{r}_{i}])]\), where:

%\begin{equation} \label{eq:1}
%\Omega([\textbf{r}_{i}])\equiv \dfrac{1}{2}\sum\limits_{i\neq j=1}^N %\varphi_{0}(|\textbf{r}_{i}-\textbf{r}_{j}|)
%\end{equation}

\[\Omega([\textbf{r}_{i}])\equiv \dfrac{1}{2}\sum\limits_{i\neq j=1}^N \varphi_{0}(|\textbf{r}_{i}-\textbf{r}_{j}|) \tag{A1}\label{myeq}\]

\noindent is the total potential energy due to the non-hard-core portion of the pair potential \(\varphi_{0}(r)\). Eqn. A1 can be recast using the definition of the 3D dirac delta function \(\delta(\textbf{x})\):
%\begin{multline*}
%\begin{eqnarray} \nonumber

%\protect\\ 

\begin{widetext}

%\begin{equation} \label{eq:2}
%\Omega([\textbf{r}_{i}])\equiv \dfrac{1}{2}\sum\limits_{i\neq j=1}^N \int %d\textbf{R}_{1}\int d\textbf{R}_{2}\delta(\textbf{r}_{i}-\textbf{R}_{1})\varphi_{%0}(|\textbf{R}_{1}-\textbf{R}_{2}|)\delta(\textbf{r}_{j}-\textbf{R}_{2})
%\end{equation}

\[\Omega([\textbf{r}_{i}])\equiv \dfrac{1}{2}\sum\limits_{i\neq j=1}^N \int d\textbf{R}_{1}\int d\textbf{R}_{2}\delta(\textbf{r}_{i}-\textbf{R}_{1})\varphi_{0}(|\textbf{R}_{1}-\textbf{R}_{2}|)\delta(\textbf{r}_{j}-\textbf{R}_{2}) \tag{A2}\label{myeq}\]

\end{widetext}

%\end{eqnarray}
%\end{multline*}

\noindent Since Eqn. A2 is a convolution with respect to \(\textbf{R}_{1}\) and \(\textbf{R}_{2}\), it can be recast as a single integral in Fourier space using the Fourier transformed potential \(\omega(k)\equiv \textup{FT}[\varphi_{0}(r)]\):

%\begin{equation} \label{eq:3}
%\Omega([\textbf{r}_{i}])\equiv \dfrac{1}{2}\sum\limits_{i\neq j=1}^N \dfrac{1}{(2\pi)^3} \int d\textbf{k} e^{-i\textbf{k}\cdot\textbf{r}_{i}} \omega(k) e^{i\textbf{k}\cdot\textbf{r}_{j}}
%\end{equation}

\[\Omega([\textbf{r}_{i}])\equiv \dfrac{1}{2}\sum\limits_{i\neq j=1}^N \dfrac{1}{(2\pi)^3} \int d\textbf{k} e^{-i\textbf{k}\cdot\textbf{r}_{i}} \omega(k) e^{i\textbf{k}\cdot\textbf{r}_{j}} \tag{A3}\label{myeq}\]

Moving the sum inside the integral in Eqn. A3 and using the definition of the non-ensemble averaged total correlation function,

%\begin{equation} \label{eq:4}
%\tilde{h}(k;[\textbf{r}_{i}])\equiv (\rho N)^{-1}\sum\limits_{i\neq j=1}^N \textup{exp}[-\textup{i} \textbf{k} \cdot (\textbf{r}_{i}-\textbf{r}_{j})]
%\end{equation}

\[\tilde{h}(k;[\textbf{r}_{i}])\equiv (\rho N)^{-1}\sum\limits_{i\neq j=1}^N \textup{exp}[-\textup{i} \textbf{k} \cdot (\textbf{r}_{i}-\textbf{r}_{j})] \tag{A4}\label{myeq}\]

\noindent one can subsequently write 

%\begin{equation} \label{eq:5}
%\Omega([\textbf{r}_{i}])=\dfrac{N\rho}{2(2\pi)^{3}}\int d\textbf{k}\omega(k)\tilde{h}(k;[\textbf{r}_{i}])
%\end{equation}

\[\Omega([\textbf{r}_{i}])=\dfrac{N\rho}{2(2\pi)^{3}}\int d\textbf{k}\omega(k)\tilde{h}(k;[\textbf{r}_{i}]) \tag{A5}\label{myeq}\]

\noindent which makes explicit the role \(\omega(k)\) plays in favoring [\(\textbf{r}_{i}\)] states possessing certain oscillatory structural correlations. Namely, any thermodynamically favorable configuration \([\textbf{r}_{i}^*]\), as weighted by \(\textup{exp}[-\beta \Omega([\textbf{r}_{i}])]\), is captured by the equilibrium average total correlation function \(h(k)\approx\tilde{h}(k;[\textbf{r}_{i}^*])\). In turn, \(\omega(k)\) sets the energetic ``preference'' for configurations structured at certain wavelengths \(k\), which appear as peaks in the structure factor since \(S(k)\equiv 1+\rho h(k)\).

% ===============================================================================================================
% &&&&&
% &&&
% &

% BIBLIOGRAPHY


%\bibliography{Bibliography}

\end{document}